\documentclass{aa} 
\usepackage{graphicx}
\usepackage{txfonts}
\usepackage{textcomp}

\begin{document}

\title{Face to phase with RU Lupi
          \thanks{Based
	  on observations collected at the European Southern Observatory (65.I-0404, 69.C-0481 and 75.C-0292), the Cerro Tololo Inter-American Observatory, Chile, the NASA Infrared Telescope Facility, Hawaii, and the XMM-Newton satellite.}
	  }


   \author{G. F. Gahm\inst{1}
           \and
           H. C. Stempels\inst{2}
	   \and
	   F. M. Walter\inst{3}
	   \and
	  P. P. Petrov\inst{4}
	   \and
	   G. J. Herczeg\inst{5}
	   }

   \offprints{G. F. Gahm}

   \institute{ Stockholm Observatory, AlbaNova University Centre,
	      SE-10691 Stockholm, Sweden,\\
              email: \mbox{gahm@astro.su.se}
               \and
	     Department of Physics and Astronomy, Uppsala University, Box 516, SE-75120 Uppsala, Sweden
	      \and
	      Department of Physics and Astronomy, Stony Brook University, Stony Brook, NY 11794-3800, USA
	      \and
	      Crimean Astrophysical Observatory, p/o Nauchny, Crimea, 98409 Ukraine
	      \and 
	     The Kavli Institute for Astronomy and Astrophysics, Peking University, Yi He Yuan Lu 5, Hai Dian Qu, Beijing 100871, P. R. China  }

   \date{}

  \abstract{}{}{}{} 
   

 \abstract
   {Some classical T Tauri stars, with intense line and continuous excess emission, show extremely complex spectral variations.  }    
   {We aim to map and interpret  the spectral variations in one such extreme T Tauri star, namely RU Lupi, and to explore how the changes are related to stellar brightness and rotational phase.  }
   {We followed the star over three observing runs, each covering a few days, collecting high-resolution optical spectra. In connection to the third run, complementary NIR spectra, multicolour photometric data, and X-ray observations were obtained.}
   {The stellar photospheric absorption line spectrum is weakened by superimposed emission, and this veiling becomes extremely high on occasion. Interpreted as a variable continuous excess emission, its contribution would amount to several times the stellar continuum brightness. However, the stellar brightness does not change much when the veiling changes, and we conclude that the veiling is dominated by narrow line emission that fills in the photospheric lines. Continuous emission, originating at the hot spot on the stellar surface, plays a dominant role only at lower degrees of veiling. 
   
   The radial velocity of narrow emission components in lines of He I vary periodically in anti-phase with the stellar velocity, reflecting the location and motion of the accretion footprint. The blue-shifted wings in He I, related to a stellar wind, are remarkably stable in equivalent width. This implies that the line flux responds directly to changes in the veiling, which in turn is related to the accretion rate close to the star. In contrast, the equivalent widths of the red-shifted wings change with rotational phase. From the pattern of variability we infer that these wings originate in accreting gas close to the star, and that the accretion funnels are bent and trail the hot spot.  
   
   The profiles of the forbidden lines of [O I] and [S II] are very stable in strength and shape over the entire observing period, and like a system of narrow, blue-shifted absorption features seen in lines of Ca II and Na I, they originate at larger distances from the star in the disk wind. Slightly blue-shifted emission components are present in the forbidden lines and might be related to a wide angle molecular disk wind proposed by others.   
     }


 \maketitle
%

\section{Introduction}
\label{subsec:intro}

Classical T Tauri stars (CTTS) are low-mass pre-main-sequence objects with intense emission line spectra. In standard models most of this emission is related to magnetospheric accretion of material from circumstellar disks. The plasma falls along magnetic field lines, and eventually energy dissipates in strong shocks in a hot spot close to the stellar surface. Most stars show continuous excess emission over the optical and ultraviolet spectral regions that is thought to originate from these shocks, whereas excess emission over the infrared spectral region originates in the cooler circumstellar disks. The optical excess, called the veiling, has the effect of reducing the depth of photospheric absorption lines. Most CTTS also show signatures of stellar winds or jets, and in addition powerful flares occur, which together with chromospheric and coronal activity produce the bulk of the strongly variable X-ray emission (see Bouvier et al. \cite{bouvier07} for a review of the properties of T Tauri stars=.
 
It is not straightforward to distinguish the different structures surrounding CTTS. It is common to associate most of the broad emission seen in the hydrogen lines and metal lines, if present, with gas falling towards  the stellar magnetic poles, while narrow emission components present in, for instance, the He {\sc i} lines would arise in warmer gas behind the impact shock, the hot spot. Inverse P Cygni profiles can be present in certain lines, and this red-shifted absorption probes gas in the accretion funnels close to the star. The distinct red-shifted absorption seen in some stars in the NIR He {\sc i} line at 10 830 \AA\ was analysed in more depth in Fischer et al. (\cite{fischer08}). These absorption components, which vary in strength, can also be present in many other lines in extreme CTTS, that is, stars with large and variable veiling and spectacular emission line spectra, such as are seen in RW Aur A (Petrov et al. \cite{petrov01}). The radial velocity of the red-shifted edge of these absorptions corresponds to the expected free-fall velocity at the stellar surface. Kurosowa \& Romanova (\cite{kurosowa13}) postulated that accretion originates from one stable and one unstable regime in the disk in order to model both periodic and stochastic variations observed in the Balmer lines, and noted that variable red-shifted absorption may set in, especially in the higher Balmer lines.

Besides the broad and narrow emission components, some stars show more extended blue-shifted emission or absorption. The blue-shifted absorption components present in many stars in the NIR He {\sc i} line have been associated with a fast moving warm stellar wind (e.g. Edwards et al. \cite{edwards06}; Kwan et al. \cite{kwan07}). This wind may also be traced as blue-shifted emission in optical He {\sc i} lines, as discussed in Edwards et al. (\cite{edwards87}) and Beristain et al. (\cite{beristain01}), who also found similar wind signatures in H$\alpha$. The complex He~{\sc i} line profiles were recently modelled by Kurosowa et al. (\cite{kurosowa11}) and Kwan \& Fischer (\cite{kwan11}). Broad blue-shifted absorption components are common also in FUV lines of neutral and once ionised elements (e.g. Ardila et al. \cite{ardila02}; Herzceg et al. \cite{herczeg06}), while lines from higher ionization states, like C IV, Si IV, and N V, only show emission; with a broad and narrow component (Ardila et al. \cite{ardila13}).

In addition to the fast moving central winds there is evidence for gas flows that are launched farther out in the circumstellar disks. For example, blue-shifted forbidden lines of [O~{\sc i}] and [S~{\sc ii}] are related to these slower moving disk winds (e.g. Edwards et al. \cite{edwards87}; Hirth et al. \cite{hirth97}). The forbidden lines are, as a rule, composed of a broad and a narrow velocity component (Hartigan et al. \cite{hartigan95}), and their origins have been discussed more recently in Ercolano \& Owen (\cite{ercolano10}), Gorti et al. (\cite{gorti11}), and Rigliaco et al. (\cite{rigliaco13}) among others. The winds contain molecular gas as well, and molecular hydrogen emission can be quite extended as found by Takami et al. (\cite{takami04}) and  Beck et al. (\cite{beck08}). This relatively warm gas ($\sim$ 2000 K) can trace shock-excited molecules in the surroundings of the remote parts of the stellar wind or jet. Moreover, Bast et al. (\cite{bast11}) and Pontoppidan et al. (\cite{pontoppidan11}) observed cool molecular gas moving at low speed, which they suggested originates in wide-angle slow disk winds. The recent study of CO emission in TTS by Brown et al. (\cite{brown13}) supports the idea that thermal molecular winds are common. These slow winds may form farther out in the disk. 

Winds in CTTS have been modelled assuming one or several components, normally including a central stellar wind that is accelerated along open dipole field lines extending from the star or from the interface between star and disk, and in addition, different types of disk winds. For a review of theoretical works see Pudritz et al. (\cite{pudritz07}), and more information on models is provided in e.g. Ferreira et al. (\cite{ferreira06}), Mohanty \& Shu (\cite{mohanty08}), Salmeron et al. (\cite{salmeron11}), Panoglou et al. (\cite{panoglou12} \v{C}emelji\'{c} et al. (\cite{cem13}), and references therein.

Regarding collecting observational data for TTS, we note that since the stars typically rotate with periods of a few days, it is desirable to collect spectroscopic observations over several consecutive nights constrain the properties and morphologies of the different circumstellar components in CTTS. This monitoring, in some cases with photometric back-up,  has been made for a number of stars (e.g.  Gahm et al. \cite{gahm95}; Hessman \& Guenther {\cite{hessman97}; Smith et al. \cite{smith97}; Smith et al. \cite{smith99}; Petrov et al. \cite{petrov01}; Alencar \& Bathala \cite{alencar02}; Gameiro et al. \cite{gameiro06}, Bouvier et al. \cite{bouvier07}; Donati et al. \cite{donati10}; Donati et al. \cite{donati11a}; Donati et al. \cite{donati11b}; Dupree et al. \cite{dupree12}). 

The veiling is quantified by the veiling factor (VF), which is defined as the ratio of an assumed excess continuum to the stellar continuum. It was demonstrated in Gahm et al. (\cite{gahm08}, Paper 1) and Petrov et al. (\cite{petrov11}) from the monitoring of six extreme CTTS that the veiling in these objects consists of two components. At low and moderate degrees of veiling the continuous emission dominates, but when the veiling increases, the continuous emission contributes less and less to the increase in VF, and instead the photospheric absorption lines get more and more filled in  by narrow emission lines. As a result, VF first responds directly to stellar brightness, but when the veiling increases further the corresponding increase in stellar brightness is much smaller than if the veiling were due to a continuous excess alone. It has been proposed that massive accretion leads to enhanced chromospheric emission as a result of changes in the temperature structure in the photosphere surrounding the hot spot. This change in atmospheric parameters has recently been confirmed by model atmosphere calculations in Dodin \& Lamzin (\cite{dodin12}) and Dodin et al. (\cite{dodin13}), although in their treatment the narrow emission lines appear even at low veiling. Recently, Alencar et al. (\cite{alencar12}) found that red wings in H$\alpha$ and H$\beta$ in V2129 Oph are strongest at a phase after the hot spot has rotated across the central stellar meridian facing us. They conclude that the accretion flow is trailing the hot spot, and these geometries, with spiral-shaped accretion funnels, were recently modelled by Romanova et al. (\cite{romanova11}). 

The general picture that has emerged from all these studies is that accretion occurs preferentially along a stellar dipole magnetic field, which may be tilted relative to the rotational axis. The hot spots at the stellar surface are in many cases surrounded by, or close to, extended dark spots, resulting in periodic radial velocity changes in the photospheric lines in anti-phase with the narrow He {\sc i} components. Accreting gas may also be redirected by multipole magnetic components to other sites on the stellar surface (e.g. Mohanty et al. \cite{mohanty08}). The details of how different spectral features vary with veiling and phase is therefore strongly dependent on the orientations of the star and magnetosphere, the accretion rate, and irregular variations can be frequent as well. 

The present study is based on data collected during the monitoring campaigns reported in Paper 1, and we focus on one extreme CTTS, namely RU Lupi, since we have found remarkably spectral variations related to accretion and winds in this object. These findings are of more general importance for understanding the nature of such active TTS and their circumstellar environments. 

The star RU Lup exhibits small-amplitude periodic ($\sim 3.7^{d}$) radial velocity variations (Stempels et al. \cite{st07}, Paper 2), and the photospheric absorption lines are narrow, $v \sin i$  of 9.0 km s$^{-1}$, indicating that the star is seen nearly pole-on (Stempels \& Piskunov \cite{st02}).} Furthermore, the broad blue-shifted absorption components flanking certain FUV emission lines indicate that we are facing RU Lup downstream in a pronounced stellar wind or jet (Lamzin \cite{lamzin00}; Herczeg et al. \cite{herczeg05}; Guenther \& Schmitt \cite{guenther08}). In agreement, Podio et al. ({\cite{podio08}) found that the NIR He~{\sc i}~line shows P Cygni absorption, and noted that the associated emission is extended, as was previously found by Takami et al. ({\cite{takami01}) for  H$\alpha$.  In addition, Herczeg et al. (\cite{herczeg05}) found that blue-shifted fluorescent H$_{2}$ emission extends smoothly from $-15$ km s$^{-1}$ to $-100$ km s$^{-1}$, and that the gas moving at around $-15$ km~s$^{-1}$ is extended. The south-west extension of the bipolar jet is aligned with the Herbig-Haro object HH 55, located 2$\arcmin$ from the star. It therefore appears that the object (both the disk and stellar axes) is slightly inclined relative to the line of sight, and in Paper 2 a stellar inclination of $i = 24\degr$ was derived.  

\section{Observations}
\label{sec:obs}

High-resolution spectra of RU Lup were obtained with the UVES spectrograph in the dichroic-mode on the \mbox{8-m} VLT/UT2 of the European Southern Observatory at Paranal, Chile. The data were collected during three observing runs, April 15\,--\,16, 2000; April 15\,--\,18, 2002; and August 14\,--\,17, 2005 (65.I-0404, 69.C-0481, and 75.C-0292). In total 29 blue and 66 red spectra were collected. The spectra have a high signal-to-noise ratio ($\ga 100$), a spectral resolution of $R \approx 60\,000$, and a wavelength coverage from 3500 to 6700 {\AA}. For more information on observations and data reduction procedures see Paper~2. Back-up spectroscopic observations were made under rather poor weather conditions during the 2005 campaign using the 4m Cerro Tololo Inter-American Observatory (CTIO) equipped with the low-resolution spectrograph RC Spec. During the August 2005 campaign NIR spectra were collected using the NASA Infrared Telescope Facility (IRTF) on Hawaii, and  RU Lup was observed on two nights. The spectra cover the wavelength range 0.8 -- 3 ${\mu}$m with a spectral resolution of $R = 2000$. 

In addition, we have compiled a long-term record of RU Lup since 2003 with $UBVRI/JHK$ photometry using the ANDICAM dual-channel imager on the SMARTS 1.3 m telescope at Cerro Tololo. Magnitudes were derived differentially to the flux of field comparison stars, which were calibrated to the photometric system in question. 

Finally, we obtained three XMM EPIC images of RU Lupi on August 8 and 17, and on September 06, 2005, that is before, during, and after the third ground-based campaign. Exposure times were $\approx 27~000$ sec. These observations have been discussed in Robrade \& Schmitt (\cite{robrade07}). 

\section{Results}
\label{sec:results}

Our VLT spectra show the rich optical emission line spectrum of RU Lup in all its pride. Lines of H, He {\sc i}, and once- and twice-ionised metals are in prominent emission, and forbidden lines of [O~{\sc i}] and [S~{\sc ii}] are in broad blue-shifted emission. Narrow absorption components are present in the blue wings of the Ca~{\sc ii} H~\&~K and Na~{\sc i} D lines, but there are no distinct inverse P~Cyg profiles present in any optical emission lines. In Paper 1 we showed how the equivalent widths (EWs) of He {\sc i} and [S~{\sc ii}] lines vary with the veiling factor VF, and in the following we will take a closer look at how different line components in various lines depend on VF and phase. For this we have re-measured the values of VF. The strong veiling washes out signatures of any TiO band absorption. However, the strongest bands in the deep-red are present but weak in one CTIO \'{e}chelle spectrum obtained, and we assign a spectral type of K5 to RU Lup. We have calculated the VFs, as described in Paper~1, assuming a surface temperature of $T_{\rm eff}$ = 4300 K, consistent with a subgiant of spectral type K5. The star spends most of its time in a state where VF varies between 2.5 and 4.0, and with occasional excursions to very high values. The total range is 1.5 -- 7.6. It should be noted that VF is well below 1.0 in most TTS. 

\subsection{The He {\sc i} lines}
\label{subsec:HeI}

The He {\sc i} 5876 \AA\ profiles are relatively stable in shape over the entire period, except that the extreme red wing varies considerably in strength. Figure~\ref{HeProf} shows two profiles obtained when the red wing was weak and strong. To define the positions and widths of the central components in the line, one narrow (NC) and one broad (BC), we used a Gaussian decomposition in three components. The component matching the extreme wings is very broad and shallow, and is used only to locate the central components. From this decomposition we defined intervals in the wings, where the contribution from the NC and BC components are negligible, and equivalent widths were measured separately for the blue and red wings over the shaded areas in the figure, corresponding to the velocity interval from $-400$ km s$^{-1}$ to $-200$ km s$^{-1}$ in the blue wing (EW1b), and from +175 km s$^{-1}$ to +350 km s$^{-1}$ in the red wing (EW1r). In addition, the equivalent width of the central emission (EW1c) was measured from $-100$ km s$^{-1}$ to +100 km s$^{-1}$  in all spectra for comparison. The He {\sc i}~6678\AA\ line profiles are similar, and the corresponding measures of the wings range from $-250$ km s$^{-1}$ to -150 km s$^{-1}$ (EW2b) 
and +100 km s$^{-1}$ to +200 km~s$^{-1}$ (EW2r).  

\begin{figure}
\centerline{\resizebox{9cm}{!}{\includegraphics{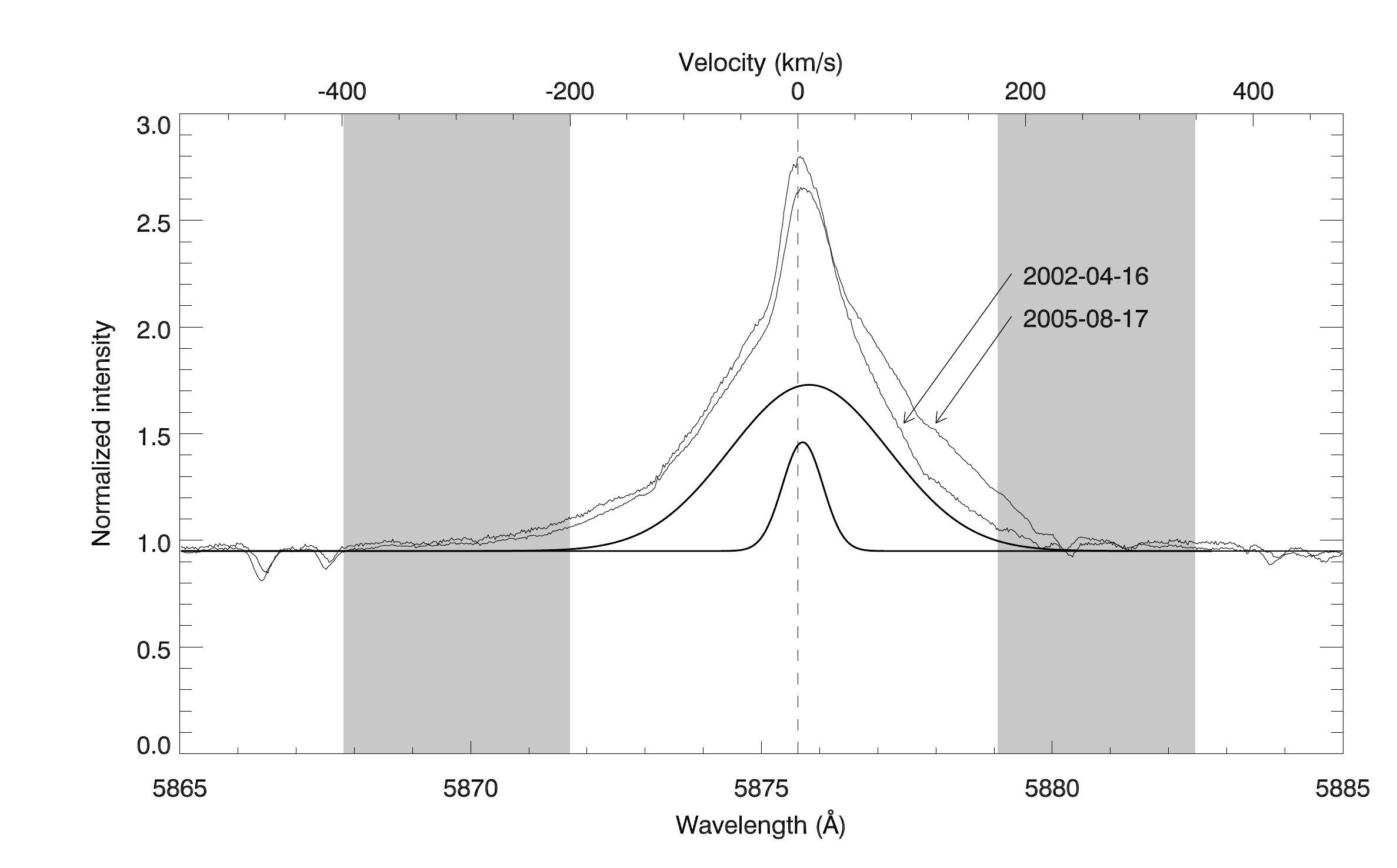}}}
\caption{ \ion{He}{i} 5876 \AA\ emission line profiles selected from two occasions when the red wing was weak (April 16, 2002) and strong (August 17, 2005). A Gaussian decomposition in three components was used to locate the two central components in all spectra, one narrow and one broad, while the third very broad and shallow component, which matches the  wings, has no physical meaning and is not depicted. Shaded areas mark the intervals in the blue and red wing not influenced by the central components, and over which equivalent widths were measured.}
\label{HeProf}
\end{figure}

\begin{figure}
\centerline{\resizebox{9.5cm}{!}{\includegraphics{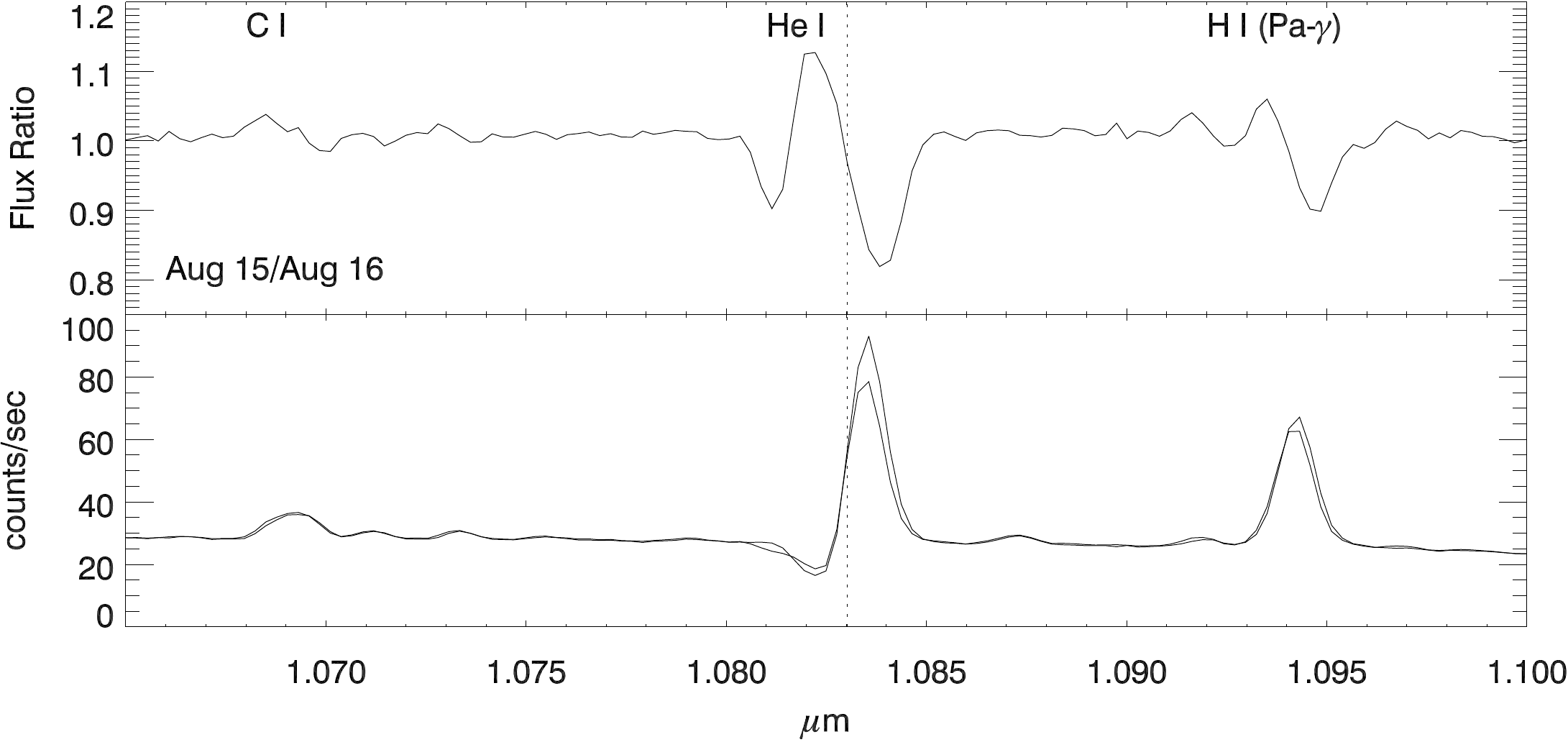}}}
\caption{The $\lambda$ 1.083 $\mu$m He~{\sc i} line on August 15 and 16, 2005. Spectra are normalized to the continuum. The upper panel shows the ratio of the spectrum of August 15 to that of August 16. The vertical line is the rest wavelength of the $\lambda$ 1.083 $\mu$m He~{\sc i} line. The emission at He~{\sc i} and at H~{\sc i} P$\gamma$ (1.094 $\mu$m) increased, while the C~{\sc i} line at 1.069 $\mu$m did not change.}
\label{NIRheI}
\end{figure}

The flux ratio of He {\sc i} 5876 \AA\ to 6678 \AA\ is sensitive to density (Shneeberger et al. \cite{schneeberger78}; Beristain et al. \cite{beristain01}), but we can only give an estimate from the ratio of EWs and correct for an energy distribution extracted from an interpolation of average SMARTS $(V - R$) colours. The interstellar extinction to RU Lup is negligible, A$_{V}$ = 0.07 (Herczeg et al. \cite{herczeg05}). We can conclude, however, that the flux ratios are very different for the blue and red wings, $\sim$~2.8 and $\sim$ 0.9, respectively. This indicates different origins of the two components, and that the density is smaller in blue-shifted gas than in red-shifted gas.

Our NIR-spectra of the He {\sc i} 10 830 \AA\ line (Fig.~\ref{NIRheI}) were obtained in 2005 during two consecutive nights over a period when the veiling dropped gradually from the extremely large values encountered on Aug. 14. There is no evidence of red-shifted absorption in any of our spectra. The profiles look similar to the one presented in Podio et al. ({\cite{podio08}) with a broad blue-shifted absorption that is more extended than the traceable blue-shifted wing in the optical He {\sc i} lines. The lines changed between August 15 and 17 in the sense that the absorption component became less extended, and narrowed from $\sim$ $-850$ to $-540$ km s$^{-1}$ at the same time as the EW of the centroid emission at $\sim$ +30 km s$^{-1}$ dropped by about 20\%.     

\subsection{The forbidden lines}
\label{subsec:forbidden}

\begin{figure}
\centerline{\resizebox{9cm}{!}{\includegraphics{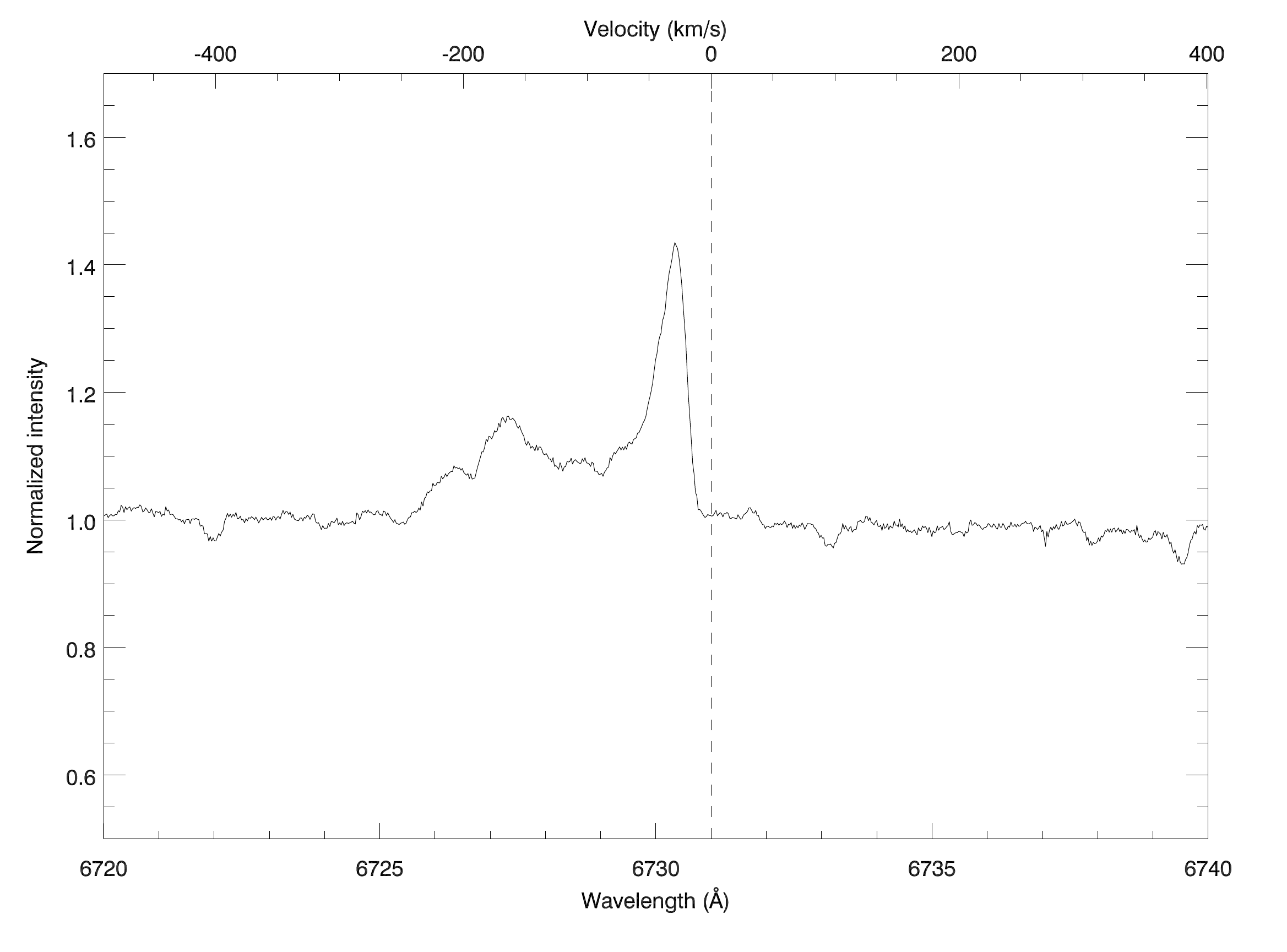}}}
\caption{Example of an [S~{\sc ii}] 6730 \AA\ emission line profile from a nightly mean from April 16, 2002}
\label{SII}
\end{figure} 

The forbidden lines of [O {\sc i}] and [S {\sc ii}] are all very similar in shape with flat, blue-shifted profiles extending to at least $-250$ km s$^{-1}$ with a narrow peak of emission at lower velocity, and a cut-off towards positive velocities. Figure~\ref{SII} shows the [S~{\sc ii}]  line at $\lambda$6730~\AA. The profiles are remarkably constant in shape over the entire observing period, and the peaks remain at a constant radial velocity (RV) of $-3.8$ $\pm$ 0.6 km s$^{-1}$ relative to the mean stellar RV. Takami et al. (\cite{takami01}) related this component to a disk wind and the broad component to the jet. The red-shifted emission from this bipolar outflow is occulted by the disk.

\subsection{Line strength and veiling}
\label{subsec:EWveiling}

\begin{figure}
\centerline{\resizebox{9cm}{!}{\includegraphics{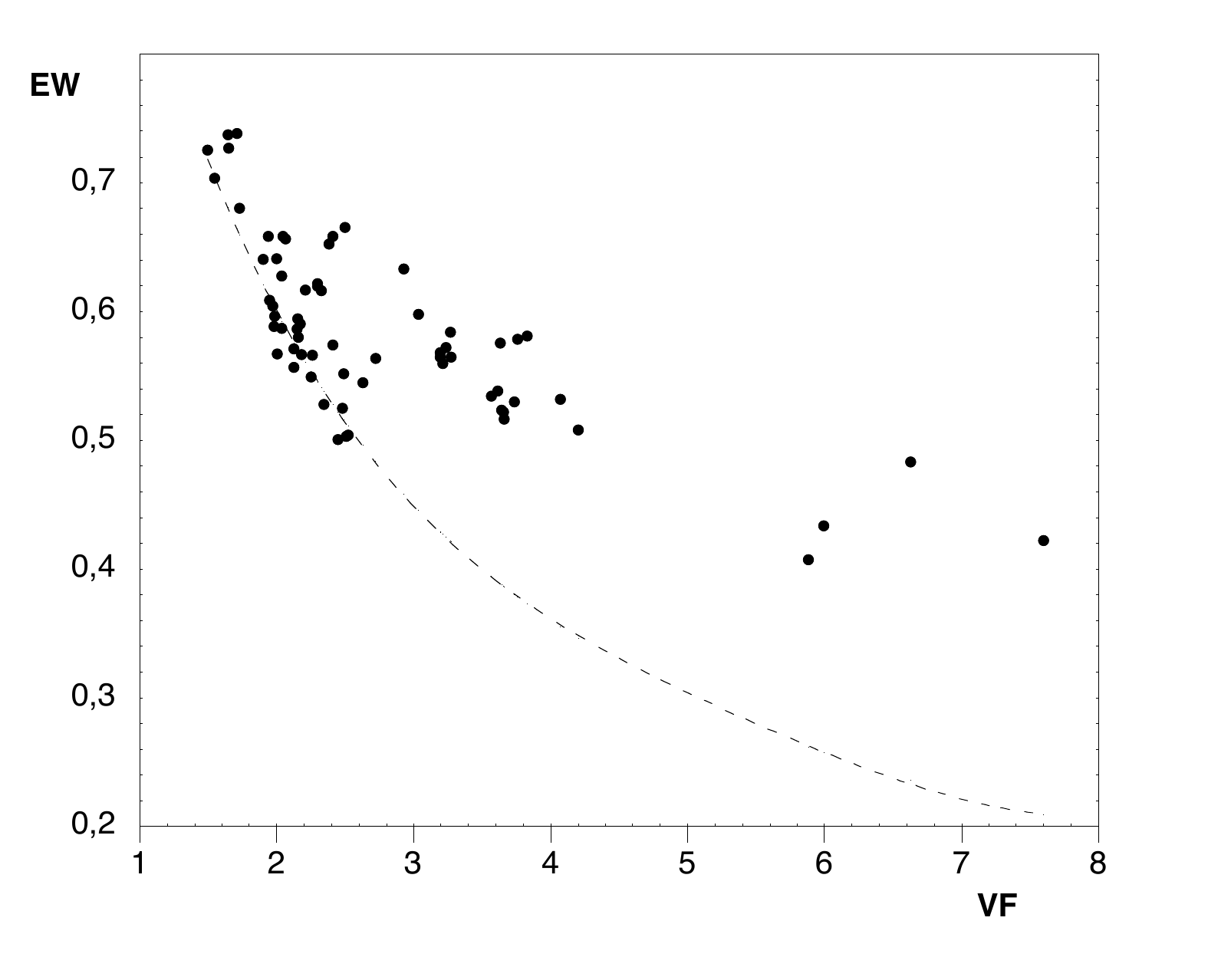}}}
\caption{The equivalent widths of [S~{\sc ii}]  6730 \AA\ versus veiling (VF). With increasing VF the values depart from what is expected, were the veiling due to a continuous excess emission alone (dashed curve). }
\label{SIIVF}
\end{figure}

In Paper 1 we demonstrated that EWs of forbidden lines in RU Lup decrease with increasing VF, but that the line flux remained constant during the first observing run for which flux calibration from standard stars could be achieved. The changes in EW for the [S~{\sc ii}]  6730 \AA\ line as observed over all three runs are shown in Fig.~\ref{SIIVF}. The dashed curve corresponds to constant line flux but variable continuous excess emission. However, when the veiling becomes larger the EWs depart from this curve and saturate, which is a result of line filling of photospheric lines as pointed out in Paper 1. Data from different observing periods are well mixed in the diagram for VF~$<$~4.5, and there is no clear sign of any secular changes in the line strength. The scatter of points for a given VF value is larger than the expected errors in individual measurements (see Sect.~\ref{subsec:EWphase}), and hence moderate intrinsic variations in the line strength are present over the three periods. 

Most emission lines, for example of neutral and once-ionised metals, follow the same pattern but with a larger scatter for a given VF, indicating that intrinsic flux variability is larger in these lines than in the forbidden lines. The central emission in the He {\sc i} and hydrogen lines also follows the same trend.  {\it However, the extended line wings behave in an entirely different way}.  

In Fig.~\ref{EWbr} we show the EWs of the blue and the red wings in He~{\sc i}~5876~\AA\ against VF for the velocity intervals defined in Section~\ref{subsec:HeI}. Points from the three observing runs are well mixed in the diagram, except that all data with VF~$>$~5 are from the third run. The strength of the blue wing is remarkably constant over the entire range of VF, in sharp contrast to all other emission lines that decline with VF. There is a hint that EW increases slightly with VF at the low end, and this impression is strengthened because the corresponding wings in the Balmer lines show the same trend. 

The red wing in He~{\sc i}~5876 is much more vivid. The scatter in EW is substantial and much larger than the uncertainties from measurements (see Sect.~\ref{subsec:EWphase}), and the red wing even vanishes on occasions. The corresponding diagram for the wings of He {\sc i} 6678 \AA\ shows the same pattern as the 5876~\AA\ line in Fig.~\ref{EWbr}. 

\begin{figure}
\centerline{\resizebox{8.3cm}{!}{\includegraphics{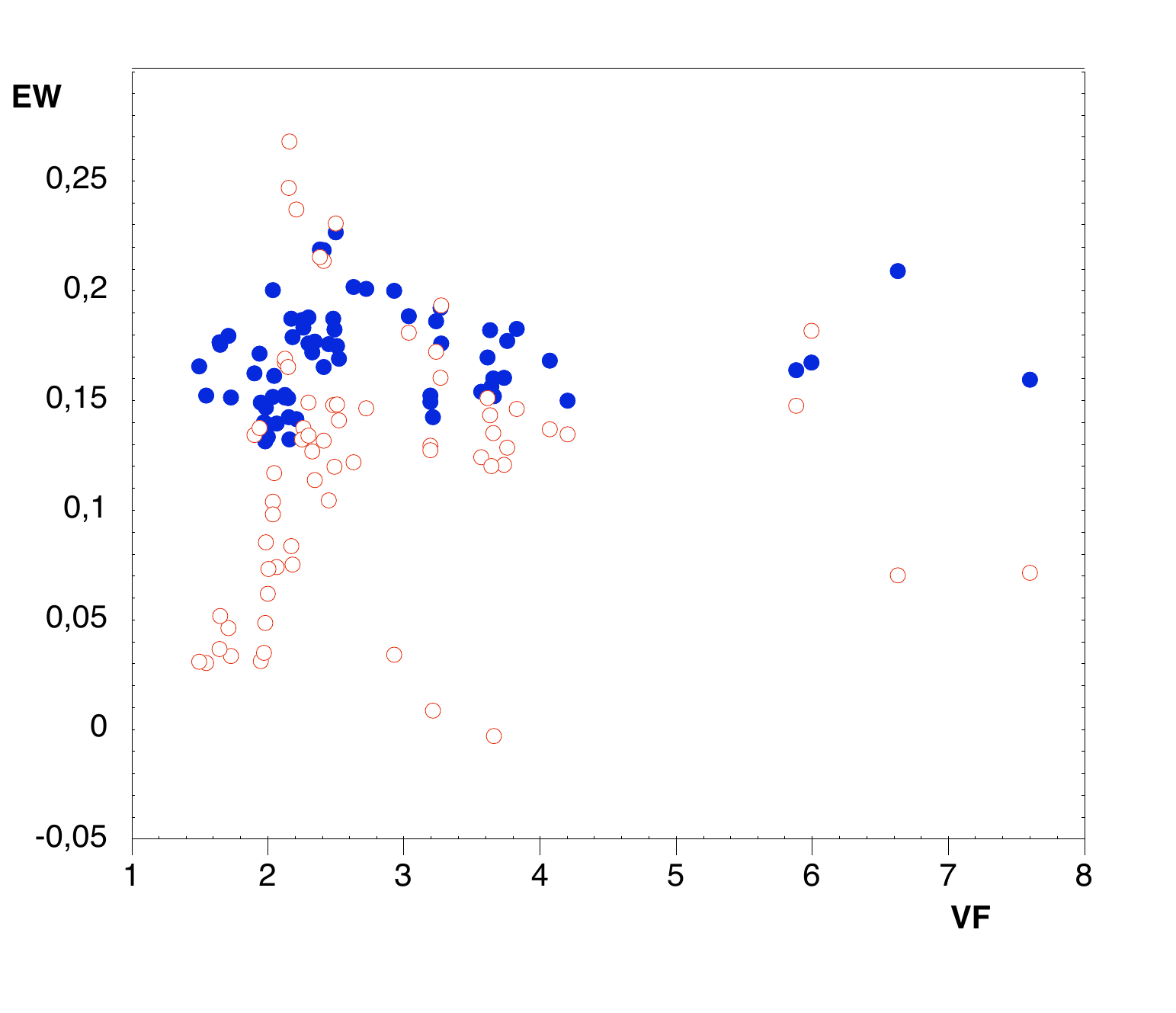}}}
\caption{Equivalent widths of the blue (filled) and red (open) wings in He~{\sc i}~5876~\AA\ versus VF.}
\label{EWbr}
\end{figure}

As shown in Paper~1, increasing veiling corresponds directly to increasing brightness of the stellar continuum for values of VF below $\approx$ 2.5, but this dependence levels off strongly when the veiling increases even more. Therefore, even when EW stays constant, the corresponding {\it line flux} increases with increasing veiling since the background continuum then increases. Since we assume that the blue wing in He~I originates in a stellar wind, we find that the flux in the wing responds directly to changes in the brightness of the spot. Therefore, the flux emitted from the wind should be directly related to the mass accretion rate. Hence, the bulk of the emission appears to come from the inner part of the warm extended blue-shifted outflow rooted close to the star. 

Our data also show that the red and blue wings are unrelated, and we identify the red-shifted emission with accreting gas, flowing towards the visible magnetic pole of the star. The large scatter in the strength of the red component could reflect erratic variations in the mass accretion rate, but additional dependencies are present, as described below. 

\subsection{Veiling, line strength, and phase}
\label{subsec:EWphase}

In Paper 2 we found that RU Lup shows persistent radial velocity variations with a period of $P = 3.71058$ days and an amplitude of $\pm$2.2 km s$^{-1}$. It was concluded that these variations reflect the rotational period of the star and result from a large spotted region. We derived a radius of $\sim 35\degr$ and a central latitude of $\sim$~60$\degr$ for this cool spotted region, and with the stellar axis inclined by $i = 24\degr$ to the line-of-sight, this region is always visible. This geometry implies that the hot spot is located inside or adjacent to the cool spot, and judging from photospheric surface maps obtained for other TTS (e.g. Donati et al. \cite{donati11a}, \cite{donati11b}) this is often the case. In the following we will investigate how various line components depend on rotational phase, and here we define phase 0.0 when the stellar radial velocity is decreasing through its mean value. At this phase the spot (including cool and hot areas) crosses the central meridian on the stellar disk from where it continues to rotate away from the observer.

To begin with, we plot RV for photospheric lines versus phase as in Paper 2, but here we also include the NC of the He {\sc i} 5876 \AA\ line (Fig.~\ref{RVs}), and we note that the NC of the He {\sc i} 6678 \AA\ line shows the same velocity fluctuations. Velocities are relative to the mean stellar heliocentric velocity, which was determined to be $-1.4$ km s$^{-1}$ in Paper~2. The diagram shows the anti-phase variations between stellar and NC velocities, as reported in Paper~1, and is similar to what has been observed in several CTTS (e.g. Petrov et al. (\cite{petrov01}) and Petrov et al. (\cite{petrov11}). 

We found no dependence of VF on phase, which is expected since the orientation of the hot spot relative to the line of sight doeas not change much. The same is true for the strength of the blue wing in the He {\sc i} lines tracing the inner part of the accelerated wind, as can be seen in the upper panel of Fig.~\ref{EWphase}. The spread in EW is very small, rms $\sim 0.02$. The formal errors from individual measurements are much smaller than this, and systematic errors must also be within the scatter present in Fig.~\ref{EWphase}. 

\begin{figure}
\centerline{\resizebox{8.7cm}{!}{\includegraphics{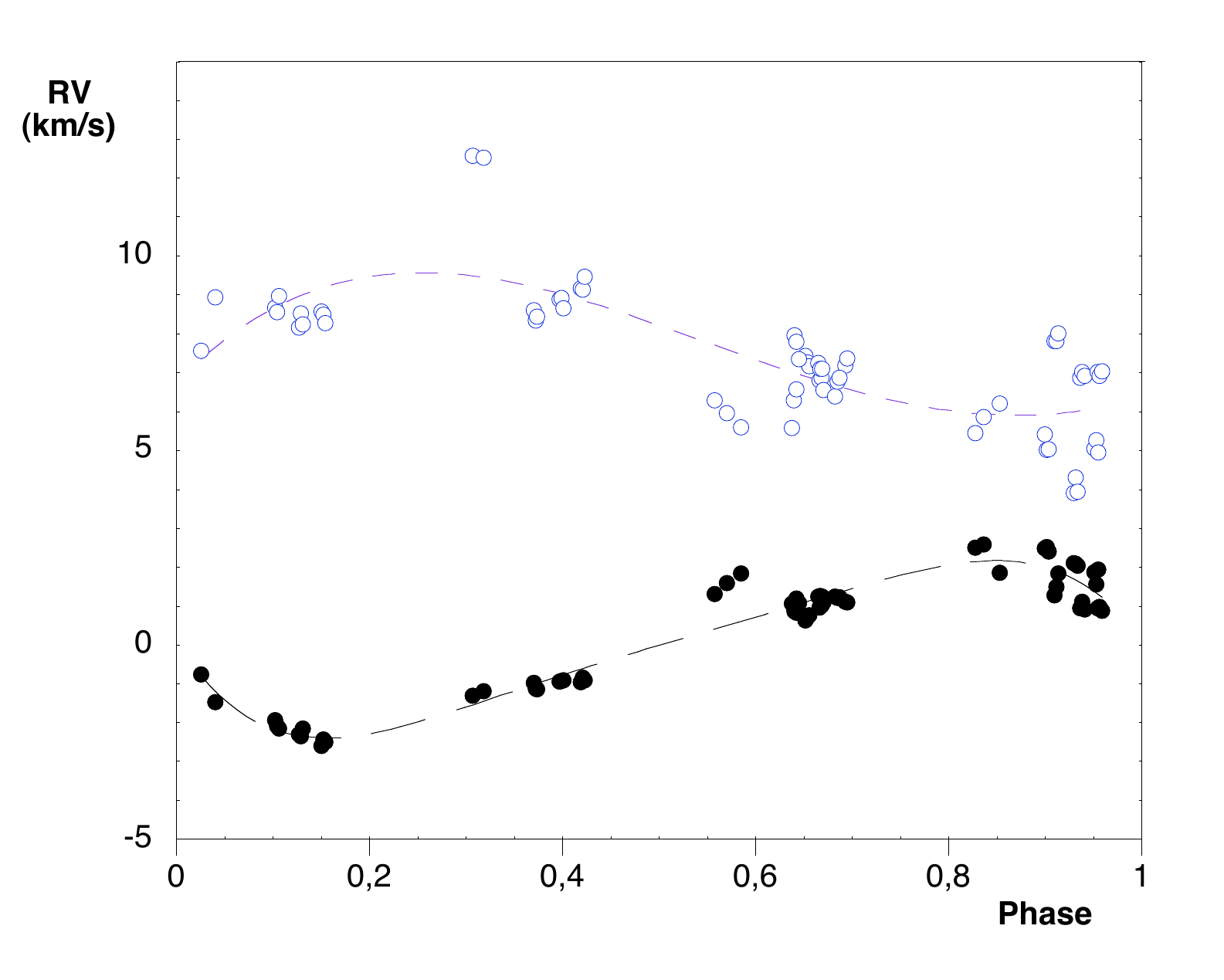}}}
\caption{Radial velocities of the star (filled) and the narrow emission components in He~{\sc i}~5876~\AA\ (open) as a function of phase. Polynomial fits to the stellar and He {\sc i} 5876 velocities are introduced. Velocities are relative to the mean stellar velocity.}
\label{RVs}
\end{figure}

\begin{figure}
\centerline{\resizebox{9cm}{!}{\includegraphics{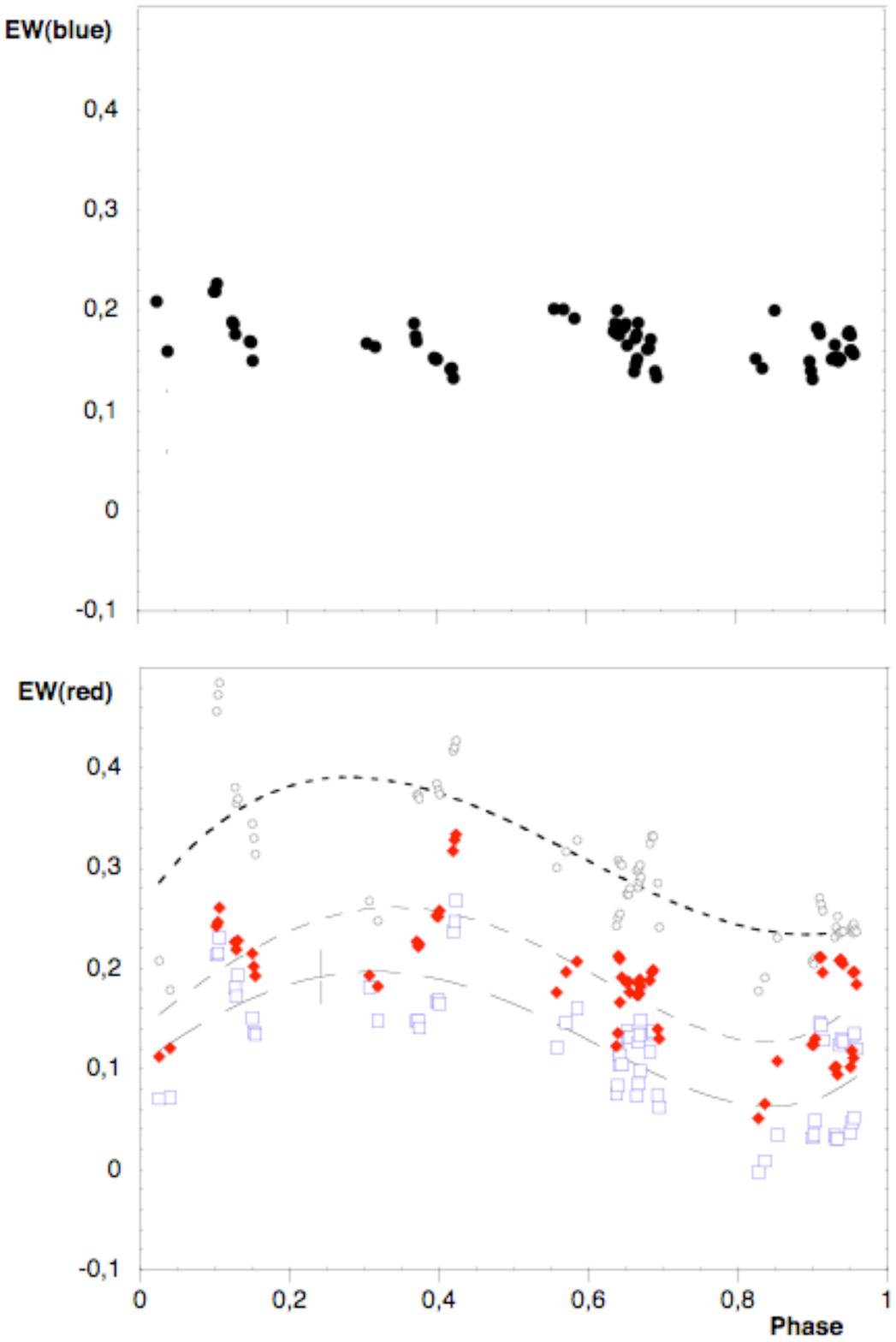}}}
\caption{Upper panel: equivalent widths of the blue wing in He~{\sc i}~5876~\AA\  as a function of rotational phase. Lower panel: equivalent widths of the red wings in He {\sc i} 5876 \AA\ (squares), He~{\sc i}~6678~\AA\ (diamonds) and H$\beta$ (open circles) as a function of phase. The EWs from H$\beta$ are divided by 10 to match the He lines. Dashed curves represent polynomial fits to the distributions.}
\label{EWphase}
\end{figure}

The red and blue wings of the He~{\sc i}~5876 line are comparable in strength, and  we expect that the intrinsic errors are the same in both cases. As can be seen in the lower panel of Fig.~\ref{EWphase}, the EWs of red wings show a much larger scatter than for  the blue wings. In addition, the strength of the red wings appears to depend on phase. The scatter is larger than the expected errors relative to the polynomial fits, but the red wings in all Balmer lines show exactly the same pattern, as shown in Fig.~\ref{EWphase} for H$\beta$ with EWs measured over the velocity interval from +175 km s$^{-1}$ to +350 km s$^{-1}$. The range in the He~{\sc i} red wing variations can be appreciated from the two extremes depicted in Fig.~\ref{HeProf}, where the spectrum with the strongest red wing was taken at phase 0.4, and the one with weakest emission at phase 0.9. We regard the fluctuations as real, and this conclusion is further supported by the fact that the blue wings in the lines in question are constant with phase, and therefore excludes any systematic effects due to the technical procedures in the data handling. In addition, points obtained from different observing runs are well mixed over the diagram. 

\begin{figure}
\centerline{\resizebox{9cm}{!}{\includegraphics{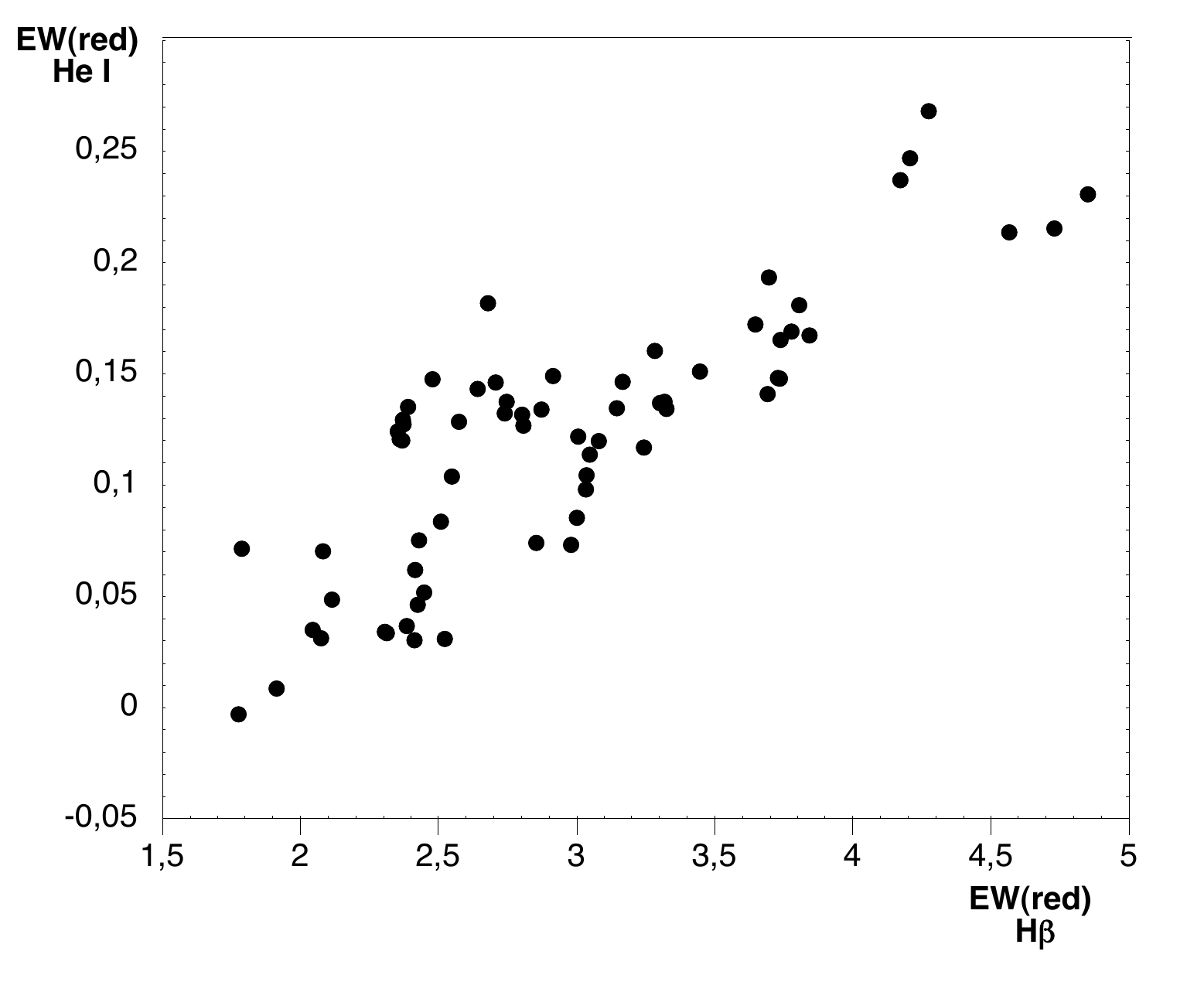}}}
\caption{Equivalent widths obtained from the red wing of He~{\sc i}~5876~\AA\ against the corresponding component in H$\beta$.}
\label{HbHe}
\end{figure}

We find that the fluctuations are most likely dominated by actual changes in the emission from accreting gas at high projected velocities in the line of sight, and not to a varying degree of red-shifted absorption in the lines. In the latter case one could expect that during an event of enhanced accretion, when the veiling increases, the opacity in the line wings increases and red-shifted absorption sets in. This would lead to decreasing emission strength in the red wings. However, EWs obtained from the He {\sc i} lines do not correlate with VF within a given interval in phase. Furthermore, red-shifted absorption is seen preferentially in the NIR He line because of its high opacity due to the metastable level involved, but no such components are seen in our spectra (Fig.~\ref{NIRheI}) taken during nights when the veiling reached extremely large values. The central broad components contribute more to the strength of the red wing in the Balmer lines than in the He {\sc i} lines. Like most emission lines, the BCs decline regularly with VF, which may explain why four points derived when VF~$>$~5 fall well below the polynomial fit, as shown for H$\beta$ in Fig.~\ref{EWphase} (phases around 0.05 and 0,3), which is not the case for the corresponding points from the He {\sc i} lines. Finally, red-shifted absorption in the optical lines would be more prominent in the He lines than in the Balmer lines. Figure~\ref{HbHe} shows that the relation between the red-shifted emission in He {\sc i} 5876 \AA\  and H$\beta$ is very tight. Red-shifted absorption is expected to affect the He line with increasing opacity in the wing, in which case the EWs for the red wing in He would fall below the linear relation when the H$\beta$ wing increases in strength. 

From these arguments we conclude that emission dominates in the wings of the selected lines, and that the amount of emitting gas, seen in the line of sight at highest projected velocities towards the star, fluctuates in anti-phase with the stellar radial velocity  as shown in Fig.~\ref{EWphase}. The scatter at a given phase is related to irregular variations in the accretion stream. The phase dependence seen in Fig.~\ref{EWphase} implies that we see maximum infalling high velocity gas approximately at or slightly after the hot spot moves away at maximum speed, and not when it is passing the central meridian of the stellar disk. We conclude that the accretion streams around RU Lup are trailing the spot. In agreement, the red EWs reaches their minimum values at approximately half a phase later, when the accretion funnel is seen at a right angle from us, and when the associated emission is velocity-shifted below the limit of +175 km s$^{-1}$.

\section{Discussion} 

The picture that emerges from the results presented in Papers 1 and 2 and in Sect.~\ref{sec:results} is that of an extreme T Tauri star seen nearly pole-on with a magnetosphere that is inclined  by $\sim 30\degr$ relative to the rotational axis. The star RU Lup shows spectral signatures of both accreting and outflowing gas. The blue and red wings of the He~{\sc i} and Balmer lines vary in an uncorrelated way, and cannot be associated with supersonic turbulence in the post-shock cooling region as is suggested to be the case in TW Hya in Dupree et al. (\cite{dupree12}). Below we will discuss further the spectral signatures of winds and accretion flows in RU Lup. The veiling factor traces continuous emission from one hot spot on the visible surface, but in addition accretion-powered narrow emission lines that dominates when the veiling becomes large. During our third observing run the veiling reached unusually large values, and at the end of the section we will take a closer look at our spectroscopic and multiwavelength photometric data collected during this period.

\subsection{Wind components}
\label{subsec:winds}

\begin{figure*}
\centerline{\resizebox{16cm}{!}{\includegraphics{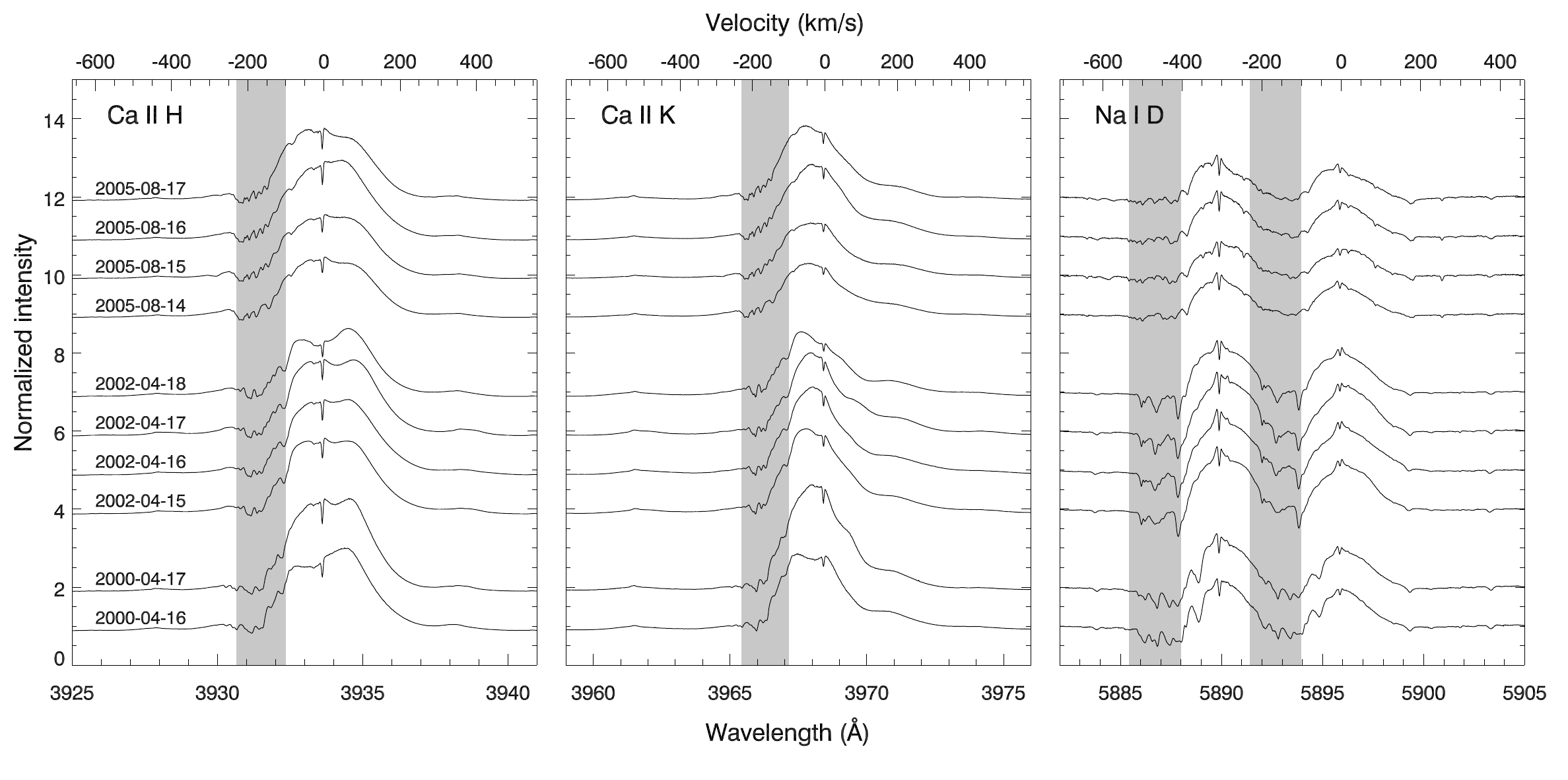}}}
\caption{Nightly means of Ca {\sc ii} H~\&~K and Na {\sc i} D line profiles ordered in time from top to bottom over the three observing periods. Narrow blue-shifted absorption components are present in the shaded intervals of all lines. The central component is interstellar. }
\label{abs}
\end{figure*}

We can distinguish two or possibly three different wind components in our spectra of RU Lup. 

1. A central warm stellar wind manifested in the blue wings of the He {\sc i} lines. These wings can be traced to $\sim$ $-400$ km s$^{-1}$, similarly for all spectra taken over all periods. The flux of the blue wings depends on the degree of veiling, which in turn depends on the accretion rate. The degree of veiling could change by more than 20$\%$ within one hour, while changes larger than 100$\%$ could occur from one night to the next, but not during a single night. 

The strong coupling between the intensities of the wind emission and that of the hot spot shows that the blue-shifted He~{\sc i} emission originates in the inner part of a central stellar wind. Whether this wind is launched along open magnetic field lines rooted at the stellar magnetic poles, or launched farther out at the interface between the magnetosphere and the disk, as in the so-called X-wind model by Shu et al. (\cite{shu94}), cannot be established from the observed time scales for changes in VF. This stellar wind component is also manifested in other lines, like the Balmer lines and the blue-shifted absorption components in the FUV Mg~{\sc ii} lines, which extend to at least -350 km s$^{-1}$ (Ardila et al. \cite{ardila02}). Moreover, the blue-shifted [Ne {\sc ii}] emission at 12.8~$\mu$m peaks at  $\sim$~$-170$ km s$^{-1}$, and was associated with a magnetically driven outflow by Sacco et al. (\cite{sacco12}). Their [Ne~{\sc ii}] line profile, which lacks a central broad component, outlines the intensity/velocity distribution of outflowing gas of low density. Emission can be traced from $\sim$~$-50$ km s$^{-1}$ to $-270$ km s$^{-1}$, but is absent or drowned in noise at higher negative velocities, where the weak wing of the He {\sc i} is still traceable in our spectra. 

2. A slower moving disk wind manifested in the broad blue-shifted emission in forbidden lines. This wind component has a cut off at velocities $\sim$ $-240$ km s$^{-1}$. Since the intensities and line profiles of these lines stay relatively constant with time, and are independent of activities close to the star, this gas flow originates far out from the star and encloses the stellar wind.

As mentioned in  Sect.~\ref{sec:results}  there are narrow absorption components in the broad emission lines of Ca~{\sc ii} H~\&~K and Na~{\sc i} D. Fig.~\ref{abs} shows nightly averages of these profiles from different observing runs. The narrow absorption components  have the same velocities both in Ca~{\sc ii} and Na~{\sc i} D, and they appear over the same velocity interval as the broad components of the forbidden lines. As can be seen in the figure, there are no significant changes that take place during one observing run, but the absorption line spectrum does change from period to period. Similar features are present in other CTTS as noted earlier by Mundt (\cite{mundt84}). The blue-shifted absorption spectrum could be related to irregularities in the disk wind, if pockets or shells of higher density form and are projected against a background of the blue wing of the Ca~{\sc ii} and Na~{\sc i} emission. These wings never extend beyond about $-250$ km s$^{-1}$, and might form in the disk wind as well. Such relatively fast moving flows have been predicted in models of MHD ejection from disks in e.g. Garcia et al. (\cite{garcia01}) and Ferreira et al. (\cite{ferreira06}).

3. There is possible evidence of a third wind component associated with the slightly blue-shifted peak in the profiles of the forbidden lines (see Fig.~\ref{SII}). These components can arise in wide-angle slow moving winds from disks seen nearly face-on as demonstrated in e.g. Appenzeller et al. (\cite{appenzeller84}) and Edwards et al. (\cite{edwards87}). For instance, the narrow peak in the [S~{\sc ii}]  6730 \AA\ line presented in Sect.~\ref{subsec:forbidden} has a heliocentric radial velocity of $-5.3$ km s$^{-1}$ and a halfwidth of 25 km s$^{-1}$. These values are very similar to those derived by Bast et al. (\cite{bast11}) and Pontoppidan et al. (\cite{pontoppidan11}) for the narrow component in the CO emission in RU Lup, $-3.9$ km s$^{-1}$ and 24 km s$^{-1}$. They interpreted their findings as evidence of a wide-angle slow-moving molecular disk wind and discussed whether it could be part of a warmer [Ne~{\sc ii}]-emitting disk wind as modelled by Alexander (\cite{alexander08}) assuming photoevaporation from the disk generated by irradiation of EUV and X-rays. However, this scenario is excluded in the case of RU Lup since the observed substantially blue-shifted [Ne~{\sc ii}] emission comes from the stellar wind, as discussed above. In addition, the velocity and width of the low-velocity component in optical [O~{\sc i}] lines were measured to -6.4 km s$^{-1}$ and 21.3km~s$^{-1}$, respectively, by Rigliaco et al. (\cite{rigliaco13}), very similar to the corresponding measures of the molecular lines components discussed above. 

Thermal photoevaporating winds driven by high-energy photons have been favoured in several studies (e.g. Gorti et al. \cite{gorti11}, Owen et al. \cite{owen10}). Rigliaco et al. (\cite{rigliaco13}), on the other hand, relates the low-velocity [O~{\sc i}] component to a photodissociation layer exposed to stellar FUV photons, and which may have a bound component in Keplerian rotation and unbound gas much farther out in the disk, where the molecular flow might form in the layer.   

To summarize, the outflowing gas from RU Lup appears to be composed of a fast central stellar wind surrounded by a disk wind moving at moderate speed. The close similarity between the line profiles of the low velocity components in [S~{\sc ii}], [O~{\sc i}], and CO strongly suggests that they are formed in the same region of a slow moving disk wind. Less certain is the indication that this gas is a part of, or an extension of, a slowly moving wide-angle disk wind containing molecules. A cut through the outflows from RU Lup could then have an onion-like structure with the slow disk wind farthest from the star. Alternatively, the slow wind forms the base of the faster disk wind, so that gas accelerates and dissociates farther out from the disk plane.

\subsection{Accretion components}
\label{subsec:accretion}

As demonstrated in Sect.~\ref{sec:results}, the strength of the red wings in the He~{\sc i} and H lines are strongly variable and change with rotational phase in a way indicating that the accretion stream is trailing the hot spot. Herczeg et al. (\cite{herczeg05}) derived a radius of 1.64 $R_{\sun}$ and a mass of 0.65 $M_{\sun}$ for RU Lup. With these values the free-fall velocity at the stellar surface is $\sim$ 390 km s$^{-1}$. The hot spot is always visible and at the phase when the red wings are strongest the hot spot is moves away at its highest positive velocity relative to the stellar mean velocity. Then, the red wing in the He {\sc i} 5876 \AA\ line can be traced to $\sim$~+360 km s$^{-1}$ and a bit farther out in the H$\delta$ line, closely matching the derived free fall velocity. The bulk of the emission in these wings originates in gas moving at  $\sim$~+150 km s$^{-1}$ along the line-of-sight, but we expect that all emission in the red wings comes from the accretion stream closest to the star. The red wing disappears half a phase later when the hot spot is moving at its highest negative velocity relative to the stellar mean velocity, and when one expects that the bulk of the emission from the accretion stream is velocity-shifted into the broad central component. However, this causes no measurable change in the strength of the BC. 

This means that the geometry of the accretion streams to RU Lup is very similar to that found for V2129 Oph by Alencar et al. (\cite{alencar12}) and which was modelled in Romanova et al. (\cite{romanova11}). In this model the accretion streams follow curved magnetic funnels that are trailing the hot spot, and it therefore appears that the streams must be anchored beyond the co-rotation radius in the disk. The fluctuations in the EWs of He~{\sc i} lines seen in Fig.~\ref{EWphase} can be reproduced in schematic geometrical models, where the bulk of the emission comes from accreting gas in trailing accretion funnels, and where matter is finally dumped radially over the hot spot, as depicted in Fig. 4 in the article by Romanova et al. (\cite{romanova11}). More detailed model simulations of these accretion geometries including line transfer calculations must be made before any conclusive comparisons between observed and modelled line profile changes in RU Lup can be made. 

\subsection{Strong activity during the third period}
\label{subsec:3rd period}

\begin{figure}
\centerline{\resizebox{9cm}{!}{\includegraphics{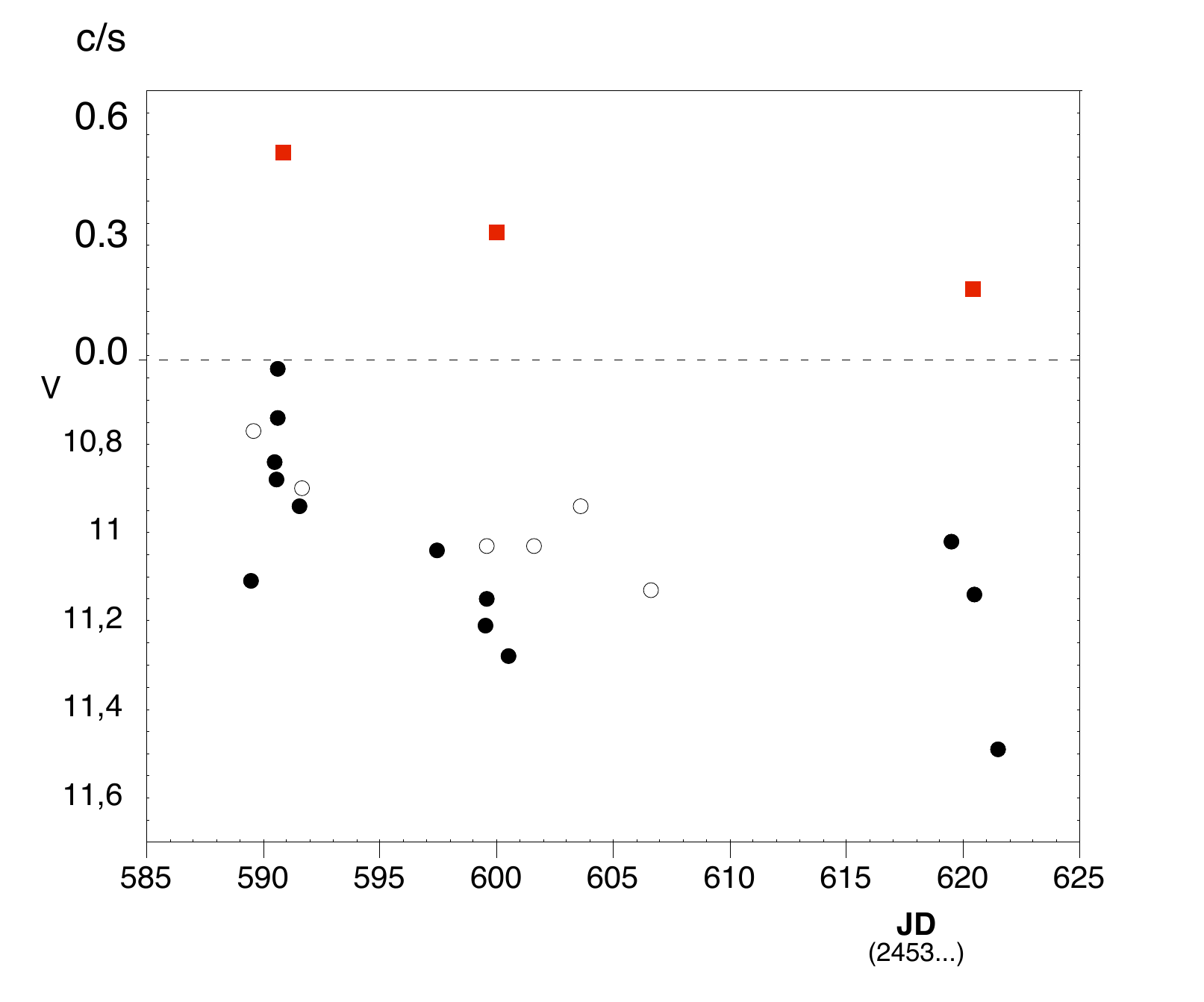}}}
\caption{X-ray flux in the energy band 0.2 - 0.5 keV  (upper panel, counts~s$^{-1}$) and $V$ magnitude against time over the third observing run. Filled circles: SMARTS data; open circles: ASAS data.}
\label{Xmag}
\end{figure}

During the 3rd observing period in 2005 the veiling reached unusually high levels. X-ray observations and broad-band photometric data were collected during this period, and also on two occasions before and after this period. As already shown in Robrade \& Schmitt (\cite{robrade07}), the star declined in X-ray brightness by a factor of $\sim~3$ from August 8 to September 6 in the 0.2 - 0.5 keV band (XMM/EPIC). A similar trend is present in our photometric data collected with SMARTS as shown in Fig.~\ref{Xmag}, and data  compiled from the All Sky Automated Survey (ASAS; see Pojamanski \cite{pojamanski02}) are added. The average decline in $V$ brightness is smaller by a factor of $\sim~1.4$, and short-term fluctuations occur, so it is difficult to judge whether there is any physical connection between the optical and X-ray variations from these data. Moreover, in the simultaneous ultraviolet photometry obtained with the optical monitor onboard the XMM the decline is less evident (see Robrade \& Schmitt \cite{robrade07}). From compiled photometric data on RU Lup we have found a mean magnitude of $V = 11.2$, and in August/September 2005 RU Lup varied irregularly between $V =  10.7 - 11.5$. The $(V~-~R)$ colour correlates with brightness in the sense that the star becomes redder with decreasing brightness in agreement with earlier results (e.g. Gahm et al. \cite{gahm74}). Our spectroscopic observations fall in the time slot 597 -  601 in the diagram, and below we will take a closer look at how RU Lup behaved during this third run.

\begin{figure}
\center{\resizebox{10cm}{!}{\includegraphics{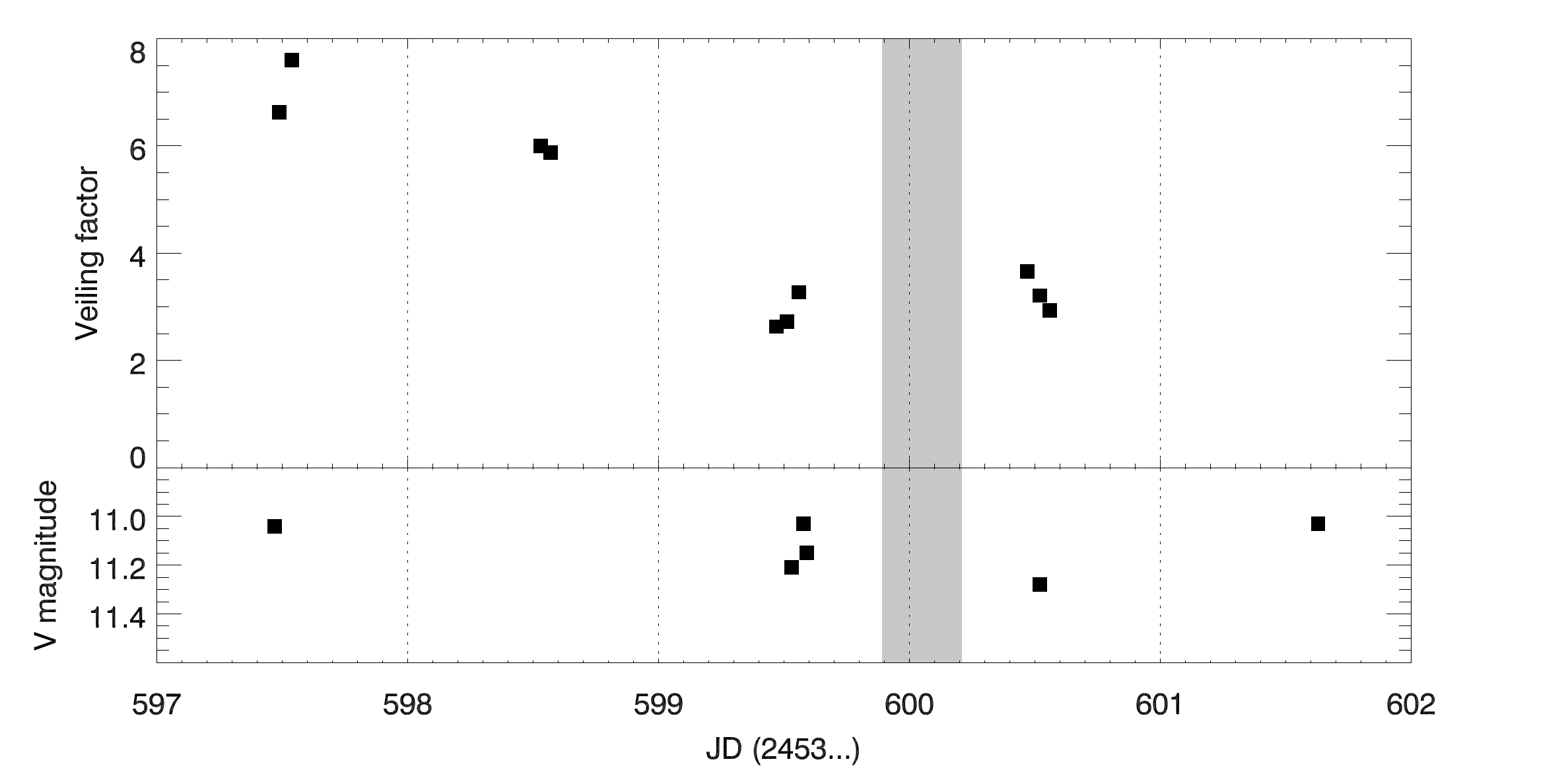}}}
\caption{Variations in veling factor (VF, upper panel) and $V$ magnitude (lower panel) over the 3rd observing period in August 2005 (photometric data from SMARTS and ASAS). The shaded area indicates the time interval over which X-ray observations were obtained (2nd point in Fig.~\ref{Xmag}). The large drop in veiling occurred from the second to the third night of ground-based observations.}  
\label{3rdAll}
\end{figure}

Figure~\ref{3rdAll} shows how RU Lup varied in $V$ brightness and veiling factor over the period when the spectroscopic observations were collected. The X-ray observations were collected in a small gap in our observations with VLT and SMARTS, but during the 25 ksec of observations in the 0.2 - 0.5 keV band the X-ray flux varied only by 20\% around a mean of 0.3 counts s$^{-1}$.  

Already at the start of the third spectroscopic run, RU Lup was found in a state of extremely large veiling, which after two days faded to more normal levels. However, the fluctuations in $V$ magnitude are small over the entire period, $\pm$~0.15 mag, and there is no hint that the stellar brightness responded to the dramatic decline in veiling factor that occurred between day 597 and 599. The first photometric point ($V$ = 11.04) in day 597 was taken at the same time as when the exposure of  the first spectrum started. From this spectrum a veiling factor of 6.6 was derived, after which VF increased even more. On the third night, VF had dropped to 2.6, when the stellar brightness declined by at most 0.15 mag. Hence, the large change in VF was not reflected by any corresponding change in the optical brightness. Similarly, the~$(B~-~V)$ data obtained over the same period do not show any related colour changes. Also from the photon counts and signal to noise in the spectra taken before and after the drop we can conclude that there was no substantial change in stellar brightness over the event. If the change in veiling were entirely due to a change in the flux from a source of continuous emission, like a hot spot, then the expected drop in $V$ magnitude would have amounted to 1.2 magnitudes over the first two nights in Fig.~\ref{3rdAll}.

The results then support the conclusions drawn in Paper~1, namely that the dominant component in the veiling factor at states of large veiling is not a continuous excess emission but is due to a forest of narrow emission lines that fill in the photospheric absorption lines (see also Petrov et al. \cite{petrov11}). However, a small decline in stellar brightness did occur during the third period, and in all photometric bands. We infer that this decline is mainly related to a decrease in veiling caused by a decrease in continuous emission at the hot spot.

\section{Conclusions}

In this paper we present results from a spectroscopic campaign following the classical T Tauri star RU Lupi over three observing runs. Complementary NIR spectra, multicolour photometric data, and X-ray observations were collected in connection with the last run. The star and its disk are seen nearly face-on, and we view the prominent outflows from the object down-stream. 

* We demonstrate that the veiling consists of two components: a continuous excess emission, and narrow emission lines filling in the photospheric absorption lines, both directly related to conditions at and around the foot-prints of the accretion funnels. This latter component dominates when the veiling becomes very large (VF $>$ 2). At this stage, a change in veiling does not lead to a corresponding change in stellar brightness, as would happen were the veiling caused by a continuous emission alone.

* The optical He~{\sc i} emission lines are complex with a narrow and broad central component, normally related to gas at the impact shock and globally accreting gas from the circumstellar disk, respectively. As reported before, the radial velocities of the narrow component is anti-correlated with the periodic stellar radial velocity, which is explained by the location of the hot spot at the stellar surface. 

* The central He~{\sc i} emission is flanked by extended wings on both sides, and we find that these wings vary in intensity independently of each other and have different origins. The blue wing is related to a warm stellar wind, and produces a prominent blue-shifted absorption component in the NIR He~{\sc i} line. The red wing varies considerably in strength, and these fluctuations are periodic and are related to stellar rotation. From the phase-dependence of these fluctuations we infer that the accretion streams close to the star are trailing the hot spot.

* The strength and profiles of the forbidden lines of [S~{\sc ii}], for example, do not change much over the entire period. These lines originate in a disk wind, and are not affected by the activity occurring closer to the star. We also discuss whether a certain narrow emission component in these lines can be related to the slow moving molecular disk wind proposed by others. In addition, there are several narrow absorption components superimposed on the blue wings of the strong emission lines of Ca {\sc ii} and Na {\sc i} caused by pockets or shells of cooler gas presumably in the disk wind. This system of absorption lines does not change from night to night, but the pattern changes completely from one observing run to the other. 
 
\begin{acknowledgements}

This work was supported by the INTAS grant 03-51-6311, the Swedish National Space Board, and the Magnus Bergvall foundation. HCS acknowledges grant 621-2009-4153 of the Swedish Research Council. FMW thanks the Provost and the office of the Vice President for Research at Stony Brook University for enabling access to the SMARTS facilities on Cerro
Tololo, which are operated by the SMARTS consortium.

\end{acknowledgements}

\end{document}